\documentclass[preprint,number,sort&compress,3p,twocolumn]{elsarticle}
\usepackage{graphicx}
\usepackage{subcaption}
\usepackage{hyperref}
\usepackage{xcolor}
\usepackage{multirow}
\begin{document}
\title{Self-Adjusting Threshold Mechanism for Pixel Detectors}

\author[lbnl]{Timon Heim}
\ead{theim@lbl.gov}
\author[lbnl]{Maurice Garcia-Sciveres}
\ead{mgs@lbl.gov}

\address[lbnl]{Lawrence Berkeley National Lab, Berkeley CA, USA}

\begin{abstract}
Readout chips of hybrid pixel detectors use a low power amplifier and threshold discrimination to process charge deposited in semiconductor sensors. Due to transistor mismatch each pixel circuit needs to be calibrated individually to achieve response uniformity. Traditionally this is addressed by programmable threshold trimming in each pixel, but requires robustness against radiation effects, temperature, and time. In this paper a self-adjusting threshold mechanism is presented, which corrects the threshold for both spatial inequality and time variation and maintains a constant response. It exploits the electrical noise as relative measure for the threshold and automatically adjust the threshold of each pixel to always achieve a uniform frequency of noise hits.
A digital implementation of the method in the form of an up/down counter and combinatorial logic filter is presented. The behavior of this circuit has been simulated to evaluate its performance and compare it to traditional calibration results. The simulation results show that this mechanism can perform equally well, but eliminates instability over time and is immune to single event upsets.
\end{abstract}

\maketitle

\section{Introduction}
\begin{figure}[htbp]
	\centering
	\includegraphics[width=0.45\textwidth]{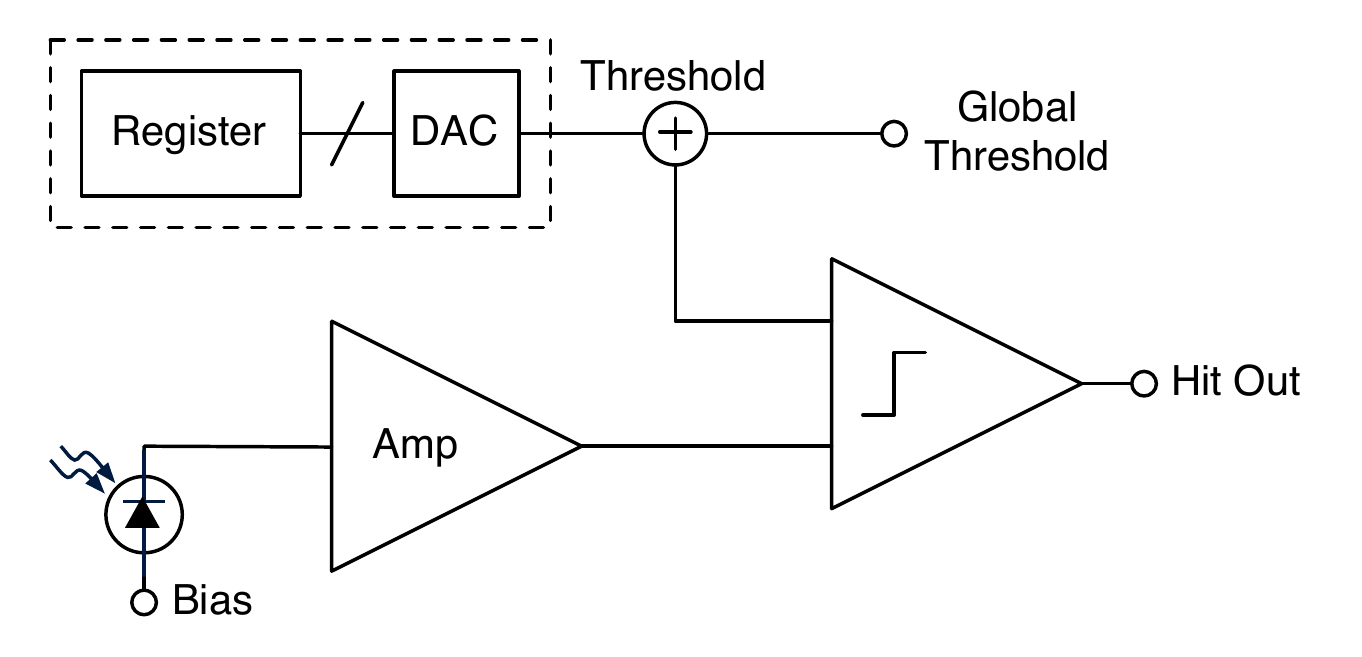}
	\caption{Schematic of generalized pixel amplifier and threshold circuit. In the following the modification of the blocks in the dashed rectangle is discussed.}
	\label{fig:pixel}
\end{figure}

An ideal pixel detector has uniform, time-invariant response across all pixels. For hybrid pixel detectors, where every pixel has its own amplifier and threshold discrimination (as shown in Fig.~\ref{fig:pixel}), uniformity is difficult to achieve due to mismatch in nominally identical CMOS transistors. This problem has been traditionally addressed by programmable threshold trimming in each pixel \cite{fei3peric, proto_fei4}. This results in spatial uniformity after trimming, but does not address time variation. Additionally, it introduces the problem of having to store configuration data in the pixels that must itself be robust against change. In a high radiation environment, stability with time (or integrated dose) and integrity of memory bits present major challenges.

A real-time measurement of the pixel threshold would permit self-trimming pixels that dynamically adjust their threshold to maintain a constant response, thus correcting for both space and time variation. 
Such a device would have advantages over existing solutions in a high radiation environment.

If the threshold is low enough, a pixel will fire randomly, at some average rate, due to noise at the amplifier output. The average firing rate rises exponentially as the threshold is decreased from a high value~\cite{diss_weingarten}, until the rate is limited by circuit bandwidth. 
In most amplifier designs, output noise voltage is dominated by the characteristics of a large input transistor. Because the device is much wider than the feature size, the pixel-to-pixel width mismatch due to the fabrication process is small. Mismatch can affect the length as transistors are short, but noise is not sensitive to small length variations~\cite{65nm_noise}. Noise, the main vice which all designs aim to minimize, thus has the virtue of good uniformity. We therefore can exploit noise to sense the response of each pixel, and adjust threshold in real-time to approximate an ideal pixel detector.

The principle of operation is a negative feedback loop whereby every noise hit increases the pixel threshold by a small amount (thus reducing the rate of noise hits) while an external ``drive'' clock periodically decreases the threshold in small steps (thus increasing the rate of noise hits). 
The user chooses a frequency of this drive clock (common to all pixels in a chip) instead of choosing a threshold.
The threshold of each pixel automatically seeks and maintains the value that results in a noise hit rate equal to the drive clock frequency.
In the absence of signals, finding the ``noise floor" is a well known technique to establish a working threshold. 
However, for real-time threshold adjustment during operation, it is critical that only hits caused by noise participate
in the feedback loop, while hits caused by the signal of interest do not. 
This requires a filter at the comparator output of Fig.~\ref{fig:pixel}, which in-situ classifies hits as noise or signal. 
Noise hits will be used to increase the threshold, and signal hits routed for further processing, storage, or readout. 
In charged particle tracking applications, we can exploit the clustered nature of signal hits to distinguish them from noise. 
In short, isolated single pixel hits with small charge are very likely to be noise and not signal \cite{diss_weingarten}. 
This distinction does not need to be perfect and we quantify performance vs. ability to instantaneously distinguish noise from signal hits. 

\section{Implementation}
The required feedback mechanism of threshold adjustment depending on noise activity can be implemented in the digital or analog domains.
In this paper we focus on the digital implementation, because it naturally fits presently used pixel architectures and still allows 
operation with externally programmed per-pixel threshold offsets, whereas the analog implementation requires a fundamentally different approach to 
threshold trimming that cannot support programmed values. The analog implementation is outlined here for completeness, 
but not analyzed further, before proceeding to the description and analysis of the digital implementation. 

\subsection{Analog Implementation}
\begin{figure}[htbp]
	\centering
	\includegraphics[width=0.45\textwidth]{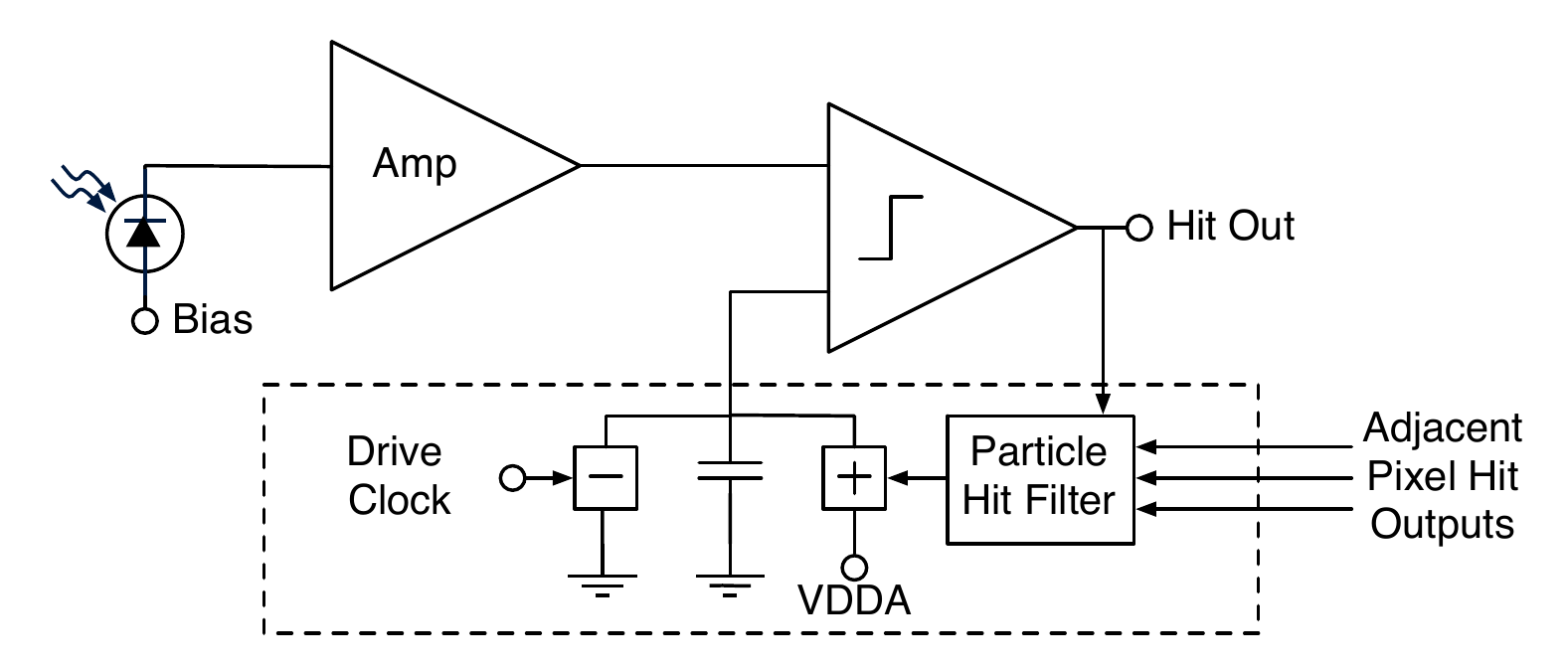}
	\caption{Analog implementation of the self-adjusting threshold mechanism.}
	\label{fig:ana_auto}
\end{figure}

The analog implementation would store a per-pixel threshold offset voltage on a capacitor fed by two charge pumps, 
one to increase the capacitor voltage and another the decrease it, as shown in Fig.~\ref{fig:ana_auto}.
Leakage currents must be low enough that the capacitor voltage variation in the time between increase/decrease events is small 
compared to the charge pump step sizes. 
Noise hits trigger the charge pump to increase the threshold, while the drive clock controls the charge pump that decreases the threshold. 
The filter to select noise hits is just as described for 
the digital implementation below. This capacitor charge pump system replaces the entire dashed box of Fig.~\ref{fig:pixel}. 
As there is no need for digital storage or for a trim DAC, and voltage storage on a capacitor has short persistence, 
there is no way to program a fixed per-pixel threshold offset.

\subsection{Digital Implementation}
\begin{figure}[htbp]
	\centering
	\includegraphics[width=0.45\textwidth]{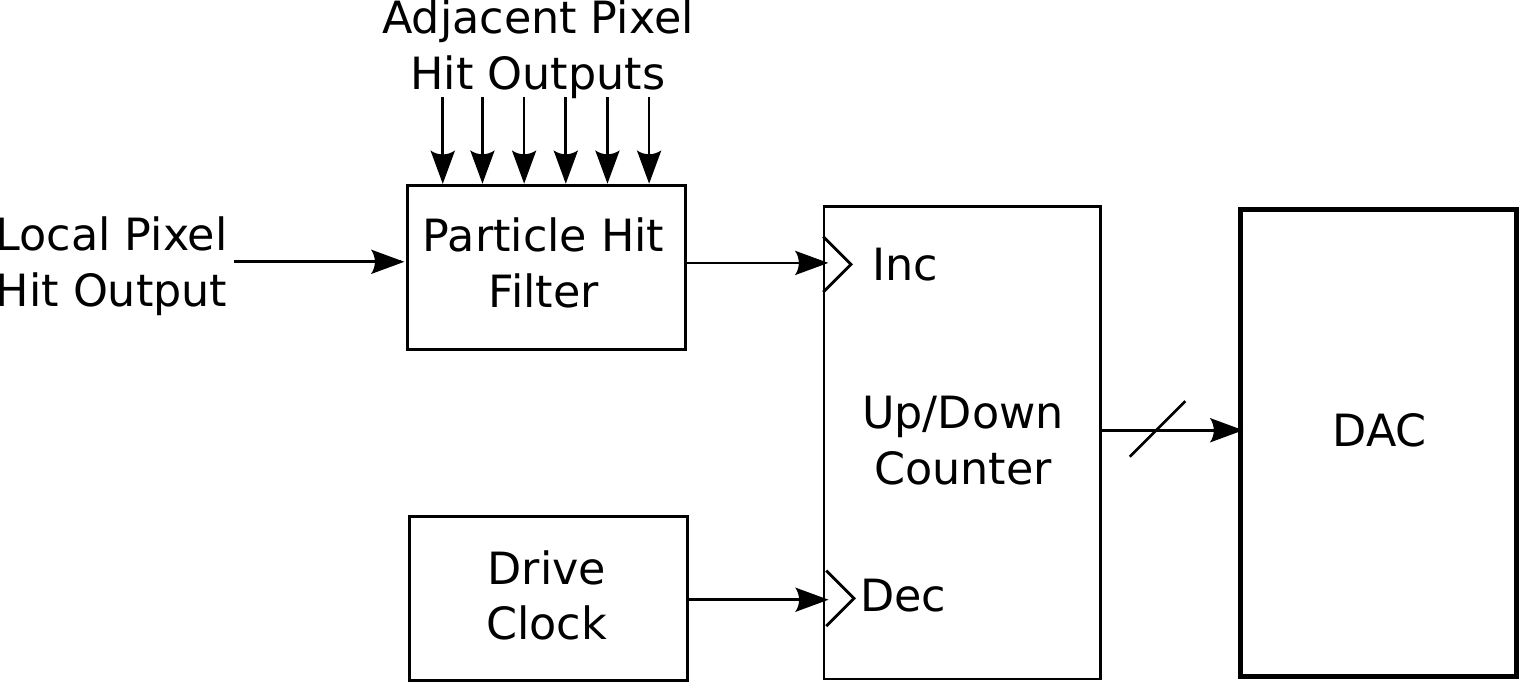}
	\caption{Block diagram of the self-adjusting threshold mechanism, which would replace the traditional trim DAC register in a pixel.}
	\label{fig:sketch}
\end{figure}

The digital implementation is an adaptation of the common architecture of Fig.~\ref{fig:pixel}. 
Instead of the static register to store a digital value, an up/down counter is used, with the counter value feeding the trim DAC, 
which is unchanged.  
A block diagram of the circuit is shown in Fig.~\ref{fig:sketch}.
The operation is similar to a tracking ADC, with the difference that it tracks frequencies instead of voltages. 
Similarly to tracking ADCs, this mechanism produces what is known as ``bit-bubble''. 
Once the system settles, the trim DAC will fluctuate between two DAC settings.
The single pixel threshold will jump up and down by one step of the DAC. Since the pixel-to-pixel
variation after classic trimming of individual pixels is also given by the trim DAC step size, this 
bit-bubble fluctuation is by construction acceptable. 

The decrement input of the up/down counter is connected to a clock source with programmable frequency (the drive clock), 
which is located on a global level and is the same for all pixels. 
The drive clock should not be synchronous for all pixels to avoid spikes in power consumption, and therefore no
clock tree is needed to distribute it. Phase dispersion can be further enhanced by adding an arbitrary delay from pixel to pixel, if needed. 
The increment input of the counter is fed by a block of combinatorial logic, which rejects hits from particles and accepts noise hits.
For pixel detectors at particle colliders, we expect particles 
to produce clusters of hit pixels (cluster size greater than 1) 
more commonly than single, isolated hits (cluster size 1)~\cite{clustersATLAS, Gjersdal2011S42}. 
This is increasingly the case as smaller and smaller pixel sizes are used.  
Furthermore, 1-pixel signal clusters will have relatively large charge, because the entire charge deposition from the source particle 
will be contained in the single pixel, whereas noise hits are by definition near threshold.  
The filter consists of am anti-coincidence logic of a pixel and its neighbors to select 1-pixel clusters, which are vetoed in case of long hits to discriminate 1-pixel particle hits from short noise hits (time is equivalent to charge).
This mechanism is used for example in the 
FE-I4 readout integrated circuit to associated small charge hits to the correct bunch crossing and therefore correct for 
time-walk~\cite{fei4hitdisc}.   
The choice of which neighbors to use will be detector-specific. The classification of hits as noise or signal does not influence which hits get stored for readout.  
Such a filter is expected to be very good at passing noise hits (so we assume 100\% to begin with and test this assumption later), 
but the particle hit rejection rate will not be perfect, 
because there can always be isolated single pixel hits caused by real signals rather than noise, 
even if such signals are not of interest for particle tracking, for example soft x-rays. One may also choose to limit the 
choice of neighbors for circuit simplicity.  
We therefore do not assume any particular rejection level for the filter, 
but analyze the performance as a function of signal hit rejection. 

If the filter rejection is 100\% (all signal hits rejected and all noise hits accepted), 
the operation of the circuit will settle and remain at a noise hit rate equal to the drive clock. 
In this case the pixel-to-pixel threshold variation is directly given by the quadratic sum of the step-size of the trim DAC and the pixel-pixel noise variation.
For particle hit rejection of less than 100\%, the particle hits passing the filter will perturb the system.
The drive clock frequency must be set higher than the rate of particle hits accepted by the filter, 
otherwise the threshold will drift to its maximum value and remain there, because the 
rate of signal hits is approximately independent of threshold (within the in-pixel adjustment range). (In most cases the threshold of a pixel at its maximum adjustment setting will still be below the charge deposited by the majority of particle hits.)
Particle hits passing the filter will also randomly increment the threshold, leading to an increased time variation (``bit bubble") of the 
threshold in each pixel (with noise alone the threshold fluctuates up and down by one DAC step). 
The threshold variation will depend on the ratio of particle hit frequency passing the filter to the drive clock frequency.
These effects lead to a trade-off between threshold equalization and noise hit rate. A toy simulation is used to analyze the performance of such a mechanism.

\section{Simulation Overview}
In the simulation a population of pixels is created. Each pixel has the following features:
\begin{itemize}
\item Base threshold: the threshold which can be increased/decreased with the trim DAC. 
\item Step size: size of a trim DAC step. Constant step size is assumed. 
\item Equivalent noise charge: the noise level which is used to generate noise hits.
\end{itemize}
All of these quantities are assumed to be Gaussian distributed for an array of pixels. The behavior of a single pixel over a long time can be simulated by creating a number of pixels all with the same settings. The width of the trim DAC of each pixel in simulation is 5~bits.

The Global variables in the simulation are:
\begin{itemize}
\item Particle hit probability: the per pixel per time unit particle hit probability.
\item Drive clock frequency: decrements the trim DAC of all pixels with a given frequency
\item Particle hit rejection: percentage of particle hits being rejected in the filter
\item Noise hit acceptance: percentage of noise hits passing the filter
\end{itemize}
As the choice of drive clock frequency is constrained by the frequency of particle hits passing the filter, only the ratio of the two is used in the following. The drive clock frequency can be derived from this ratio by scaling it to the expected particle hit frequency. The particle hit rejection is generic and does not assume any underlying model. 

A population of 1000 pixels is is used for the results presented here.
The parameters simulated are consistent with those being prototyped~\cite{RD53A} 
for detector upgrades planned for the High Luminosity Large Hadron Collider (HL-LHC).
They are listed in Table~\ref{tab:par}.

\begin{table}[htbp]
	\centering
	\caption{Default values of simulation parameters in electrons (e$^{-}$) when they are not being varied as indicated in the text.}
	\label{tab:par}
	\begin{tabular}{@{}l|ll@{}}
		\multirow{ 2}{*}{Parameter} & Distribution & \multirow{ 2}{*}{Value}\\
		 & type & \\
		\hline\hline
		Base threshold & Gaussian & 500~e$^-$ $\pm$ 100~e$^-$\\
		Step size & Gaussian & 25~e$^-$ $\pm$ 2~e$^-$\\
		ENC & Gaussian & 50~e$^-$ $\pm$ 5~e$^-$\\
		\hline
		Particle hit & \multirow{ 2}{*}{fixed} & \multirow{ 2}{*}{$10^{-3}$ per tick} \\
		probability & & \\
		Drive clock &  \multirow{ 2}{*}{fixed} &  \multirow{ 2}{*}{$10^4$ ticks}\\
		period & & \\
		Particle hit &  \multirow{ 2}{*}{fixed} &  \multirow{ 2}{*}{90\%}\\
		rejection  & & \\
		Noise hit &  \multirow{ 2}{*}{fixed} &  \multirow{ 2}{*}{100\%}\\
		acceptance & & \\
		\hline
	\end{tabular}
\end{table}

The simulation is done in discrete, equal time intervals or {\em ticks}, 
in each of which one noise or particle hit can occur (noise taking precedence over particle hits). To avoid manual tuning of the global threshold (as it would be needed in a real system to shift the mean of the base threshold distribution to the right value) in simulation, the base threshold of all pixels is adjusted automatically in steps of 50~e$^-$ if more than 10\% of all pixels have reached the maximum or minimum trim DAC value. The total numbers of noise and particle hits are recorded, as well as the pixel-to-pixel threshold variation at the end of the simulation. A total of 500,000 ticks has been identified as sufficient time for the system to stabilize. Note the effects of single pixel bit-bubble fluctuation since all pixels fluctuate out (noise is uncorrelated between pixels). 

\section{Simulation Results}
Besides testing the general functionality, the simulation can help understand two important factors of this mechanism. 
The main performance metric is the achieved threshold dispersion (including both pixel to pixel variation and time fluctuation).
It is further possible to analyze the impact of the particle hit filter rejection and set margins of 
a minimum filter performance for a given system for this mechanism to work.  

\subsection{Example of Simulation Outputs}
\begin{figure}[htbp]
	\centering
	\includegraphics[width=0.45\textwidth]{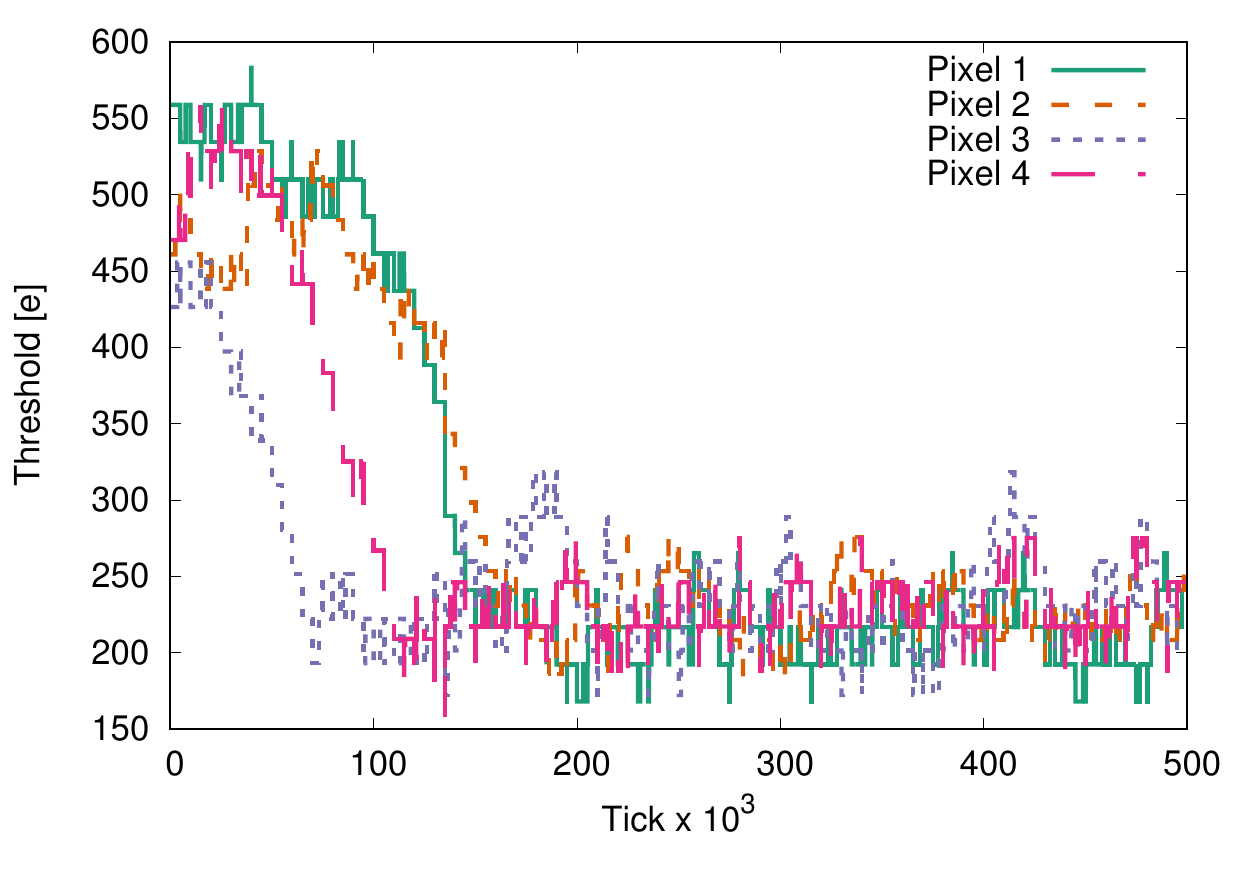}
	\caption{Simulation of the threshold over time of four randomly selected pixels over $10^6$ ticks. The simulation started with a base threshold of 500~e$^-$ $\pm$100~e$^-$, a trim DAC step size of 25~e$^-$ $\pm$2~e$^-$ and noise of 50~e$^-$ $\pm$5~e$^-$. The counter is decremented every $10^{4}$ ticks and the particle hit probability is $10^{-3}$ per tick. The particle hit rejection is 90\%.}
	\label{fig:single-pixel}
\end{figure}

\begin{figure}[htbp]
	\centering
	\includegraphics[width=0.45\textwidth]{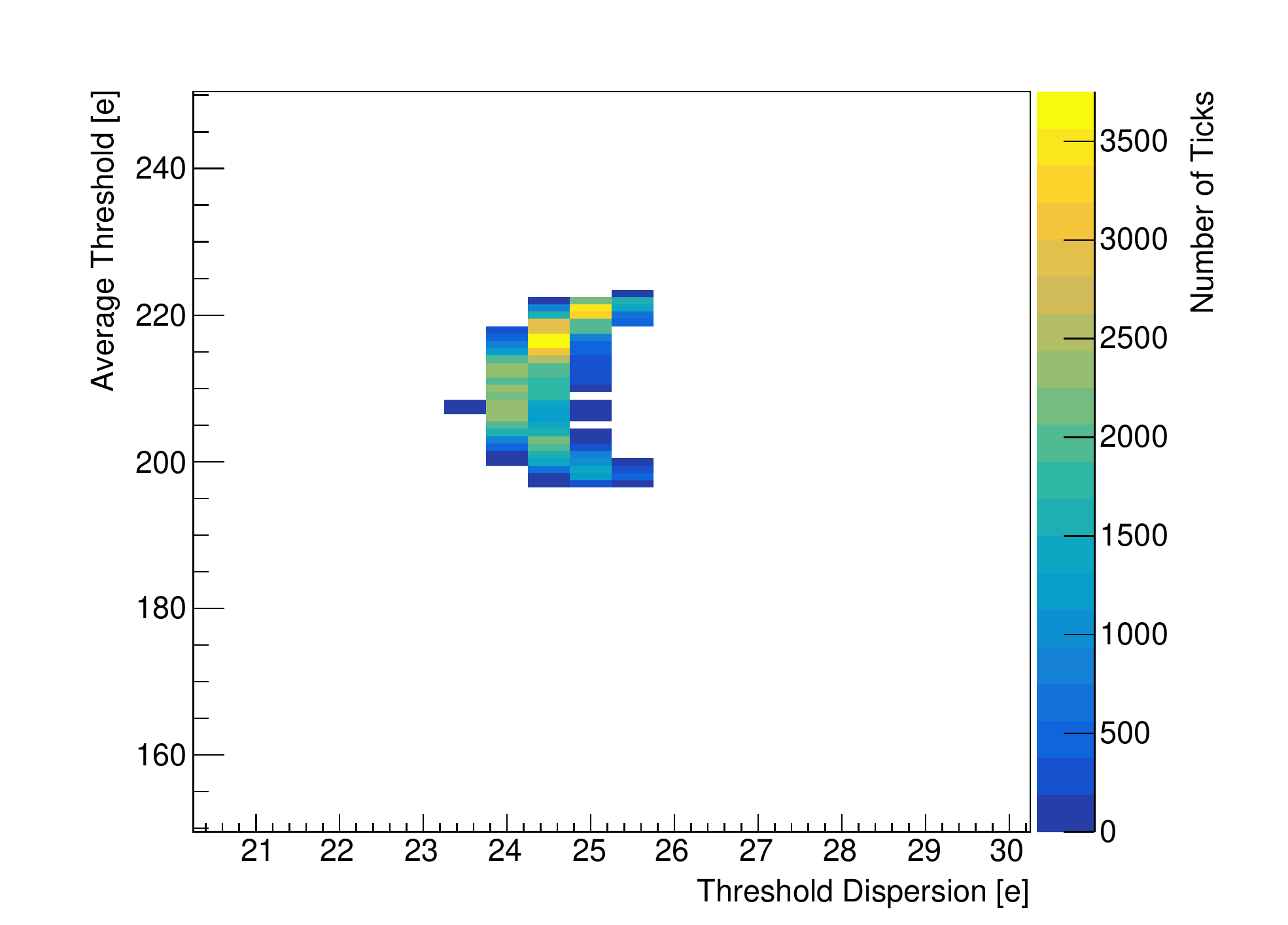}
	\caption{Average threshold and threshold dispersion 5000 pixels over the course of $100\cdot 10^{3}$~ticks. The simulation started with a base threshold of 500~e$^-$ $\pm$100~e$^-$, a trim DAC step size of 25~e$^-$ $\pm$2~e$^-$ and noise of 50~e$^-$ $\pm$5~e$^-$. The counter is decremented every $10^{4}$ ticks and the particle hit probability is $10^{-3}$ per tick. The particle hit rejection is 90\%.}
	\label{fig:thr_sigma}
\end{figure}

\begin{figure}[htbp]
	\centering
	\includegraphics[width=0.45\textwidth]{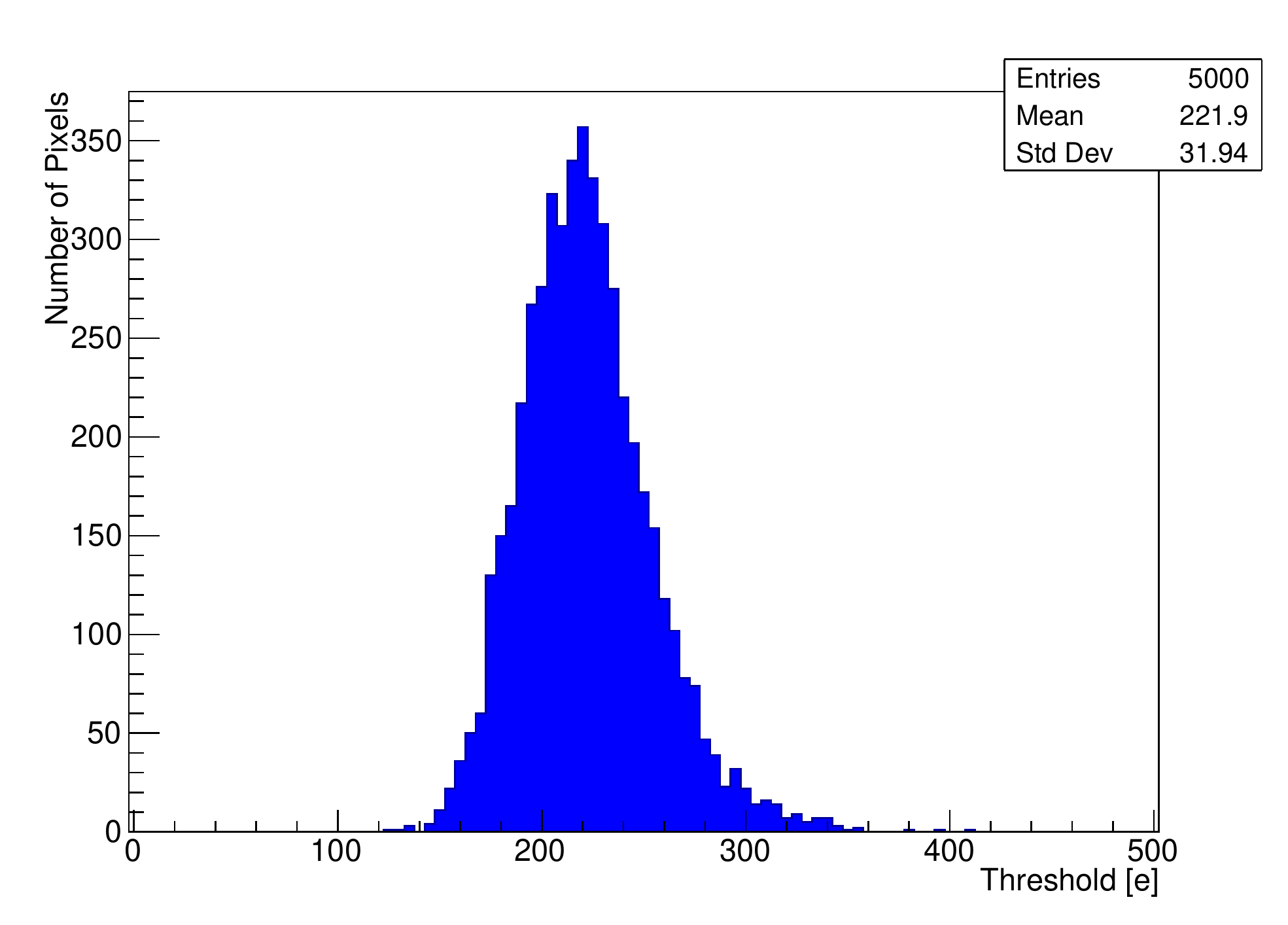}
	\caption{Threshold distribution of 5000 pixels after $500\cdot 10^{3}$~ticks. The simulation started with a base threshold of 500~e$^-$ $\pm$100~e$^-$, a trim DAC step size of 25~e$^-$ $\pm$2~e$^-$ and noise of 50~e$^-$ $\pm$5~e$^-$. The counter is decremented every $10^{4}$ ticks and the particle hit probability is $10^{-3}$ per tick. The particle hit rejection is 90\%.}
	\label{fig:threshold_example}
\end{figure}

\begin{figure}[htbp]
	\centering
	\includegraphics[width=0.45\textwidth]{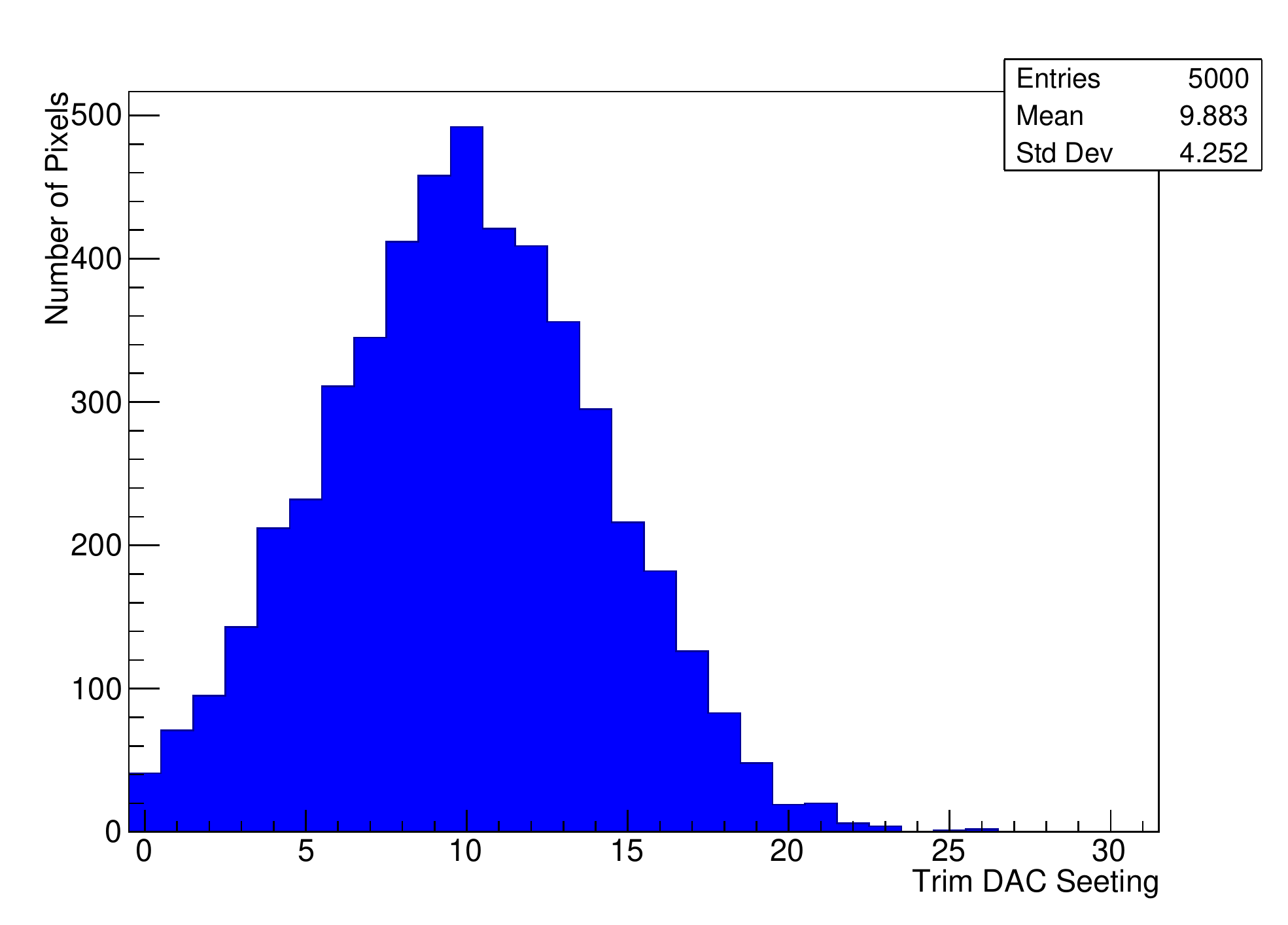}
	\caption{Trim DAC setting distribution of 5000 pixels after $500\cdot 10^{3}$~ticks. The simulation started with a base threshold of 500~e$^-$ $\pm$100~e$^-$, a trim DAC step size of 25~e$^-$ $\pm$2~e$^-$ and noise of 50~e$^-$ $\pm$5~e$^-$. The counter is decremented every $10^{4}$ ticks and the particle hit probability is $10^{-3}$ per tick. The particle hit rejection is 90\%.}
	\label{fig:tdac_example}
\end{figure}

Sample simulation outputs from 4 pixels are in Fig.~\ref{fig:single-pixel}. In this case the threshold settles around 220~e$^-$. While statistical fluctuations (bursts) of particle hits increase the threshold for a short time (for instance around $t=400\cdot 10^3$), 
it always drifts back to the optimal value to achieve the requested noise hit rate.
This behavior is also what makes the pixel single event upset immune without requiring additional logic. 
The variation over time of the average threshold and threshold dispersion is shown in Figure~\ref{fig:thr_sigma} over the course of $100\cdot 10^{3}$~ticks. The variation of the average threshold of 6.82~e$^{-}$ over time is smaller than the instantaneous threshold dispersion of around 24~e$^{-}$. Therefore, once all pixels have stabilised, the instantaneous threshold dispersion stays approximately constant.
Figure~\ref{fig:threshold_example} and \ref{fig:tdac_example} show the threshold and trim DAC setting distribution at the end of a simulation with 5000 pixels. The threshold distribution shows a tail to higher values, which can be explained by the effect of pixel thresholds being increased by particle hits. The trim DAC step size is not fully optimised as it does not use the full range, which is considered to be a realistic scenario. In the following the threshold dispersion is measured as the standard deviation at the end of the simulation. 

\subsection{Optimal Operation Conditions}
The optimal system should achieve the lowest noise hit rate (equivalent to the lowest drive clock frequency) that keeps the threshold dispersion below a certain target value. Different particle hit rejection of the filter will result in different purity of hits above threshold, defined as the ratio of signal hits over the sum of signal plus noise hits. High purity is important because the lower the purity the more readout bandwidth is needed.
For high purity one can choose to read out all hits, regardless of how they have been classified by the filter,
whereas for low purity one may need to discard the hits classified as noise and read only the non-isolated or high charge isolated hits. 
The hit purity depends on the filter signal hit rejection and the signal hit rate. 

\begin{figure}[htbp]
	\centering
	\includegraphics[width=0.45\textwidth]{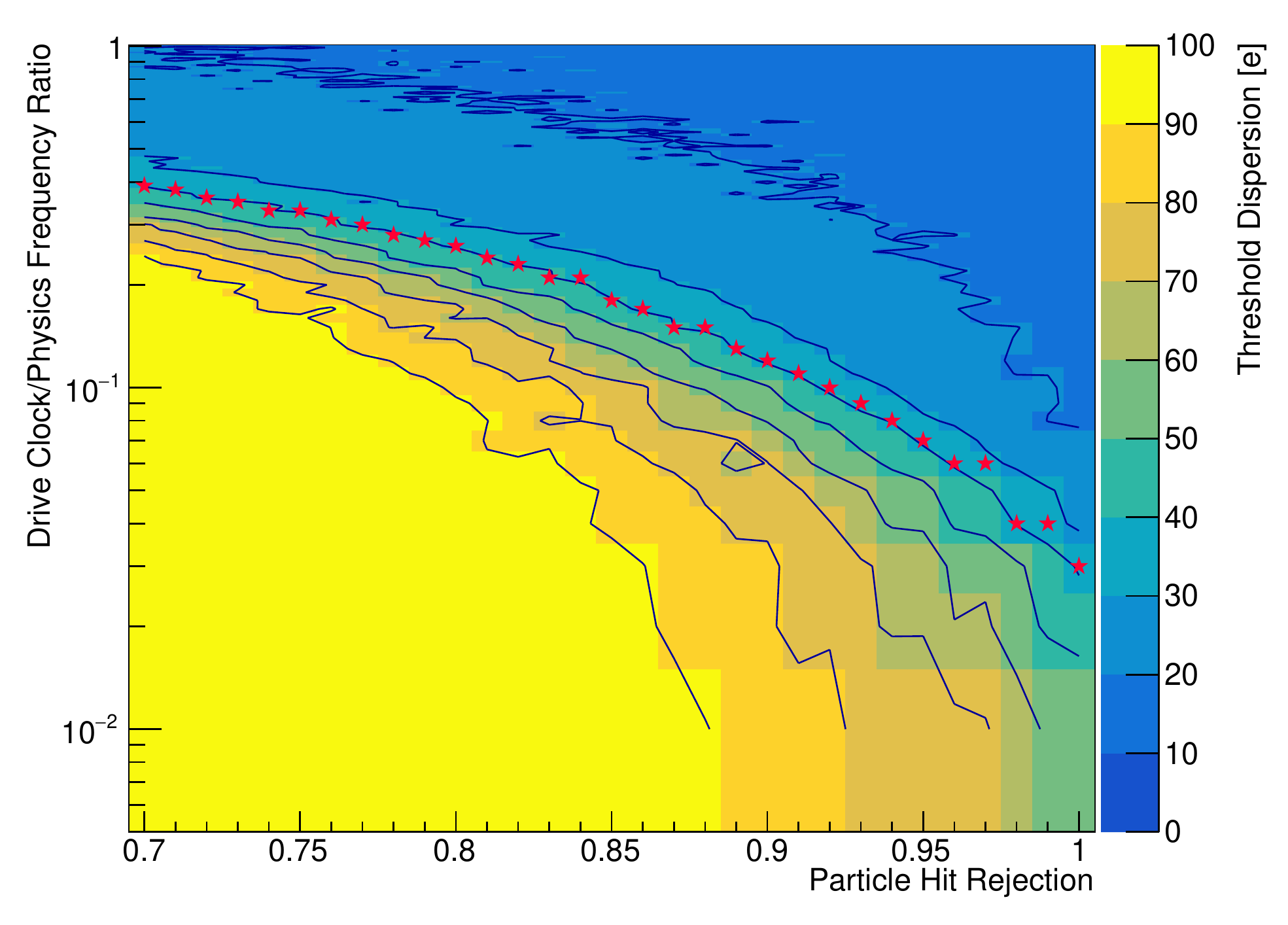}
	\caption{Threshold dispersion in dependance of normalized drive clock frequency ratio and particle hit rejection. Stars mark the lowest frequency ratio which still achieves a 40~e$^-$ threshold dispersion.}
	\label{fig:rms}
\end{figure}

\begin{figure}[htbp]
	\centering
	\includegraphics[width=0.45\textwidth]{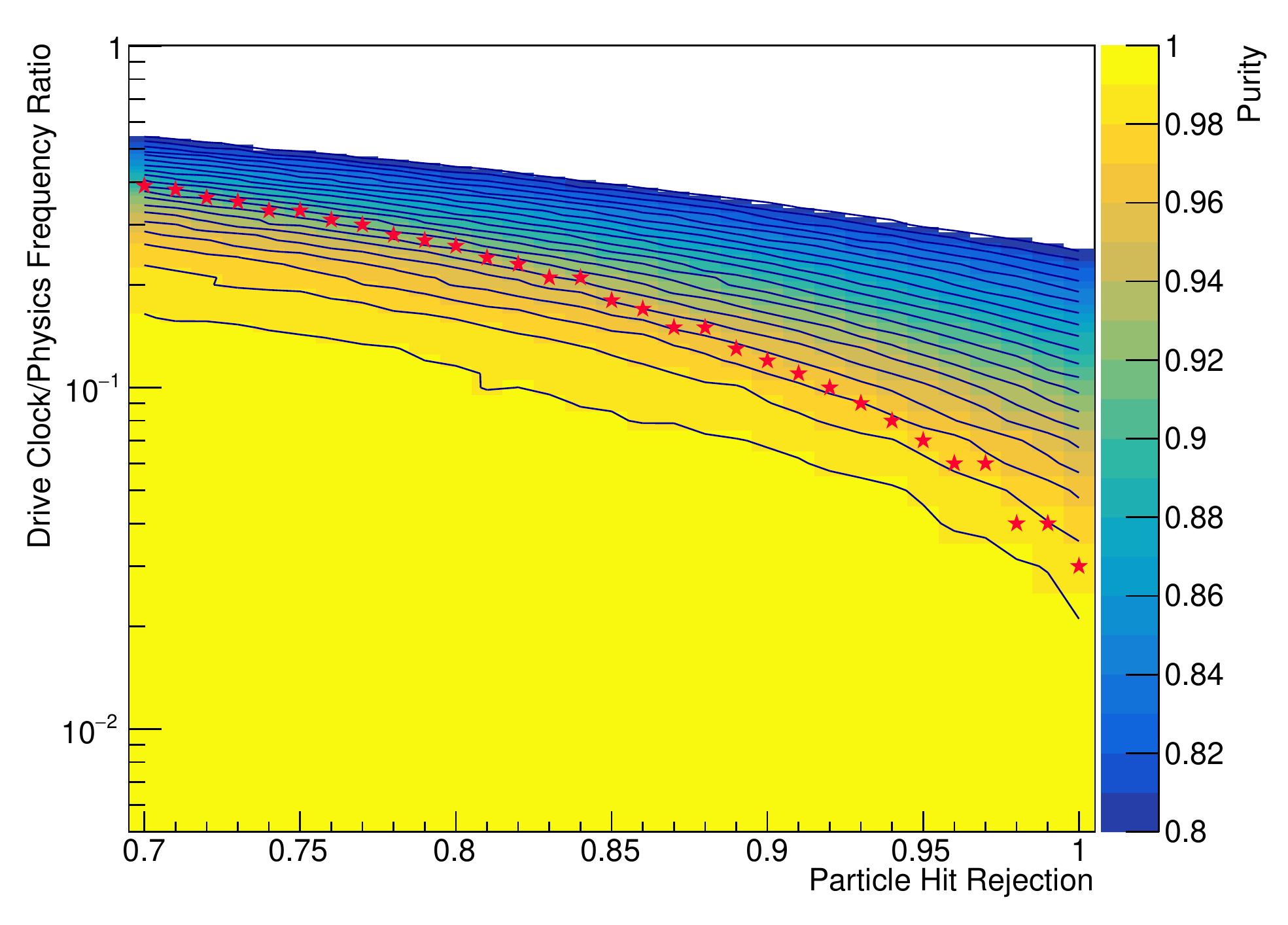}
	\caption{Purity, the ratio of signal hits over the sum of signal hits plus noise hits, in dependance of normalized drive clock frequency ratio and particle hit rejection. Stars mark the lowest frequency ratio which still achieves 40~e$^-$ threshold dispersion.}
	\label{fig:purity}
\end{figure}

Fig.~\ref{fig:rms} shows the threshold dispersion for different filter rejection and different drive clock frequency normalized to the physics hit rate. 
The drive clock frequency normalized this way naturally reflects the expected purity for a perfect filter.
If the normalized drive clock frequency is too low and the particle hit rejection is not perfect, the threshold will drift to higher and higher values. 
A higher normalized drive clock frequency will lead to good threshold dispersion, but at the expense of purity, as can be seen in Fig.~\ref{fig:purity}. We define the optimal working point as the lowest drive clock frequency at which the threshold dispersion is still below 
40~e$^-$, which is the value of threshold dispersion specified for HL-LHC prototypes and often achieved in present detectors. 
This working point is indicated by red stars in Fig.~\ref{fig:rms} and \ref{fig:purity}.
For example, the results show that a purity of 94\% can be achieved with a filter rejecting 90\% of physics hits. The normalized drive clock frequency 
for this working point is 0.16, which is slightly higher than rate of signal hits passing the filter (0.1), as expected. The contamination of 
noise hits among the total hits (1 minus the purity) is 0.06, which is in fact the value by which the drive clock frequency exceeds the 
rate of signal hits passing the filter. This makes sense, because all hits passing the filter increment the threshold, so the drive clock 
frequency must be the sum of both.     

\subsection{Noise Distribution}
\begin{figure}[htbp]
	\centering
	\includegraphics[width=0.45\textwidth]{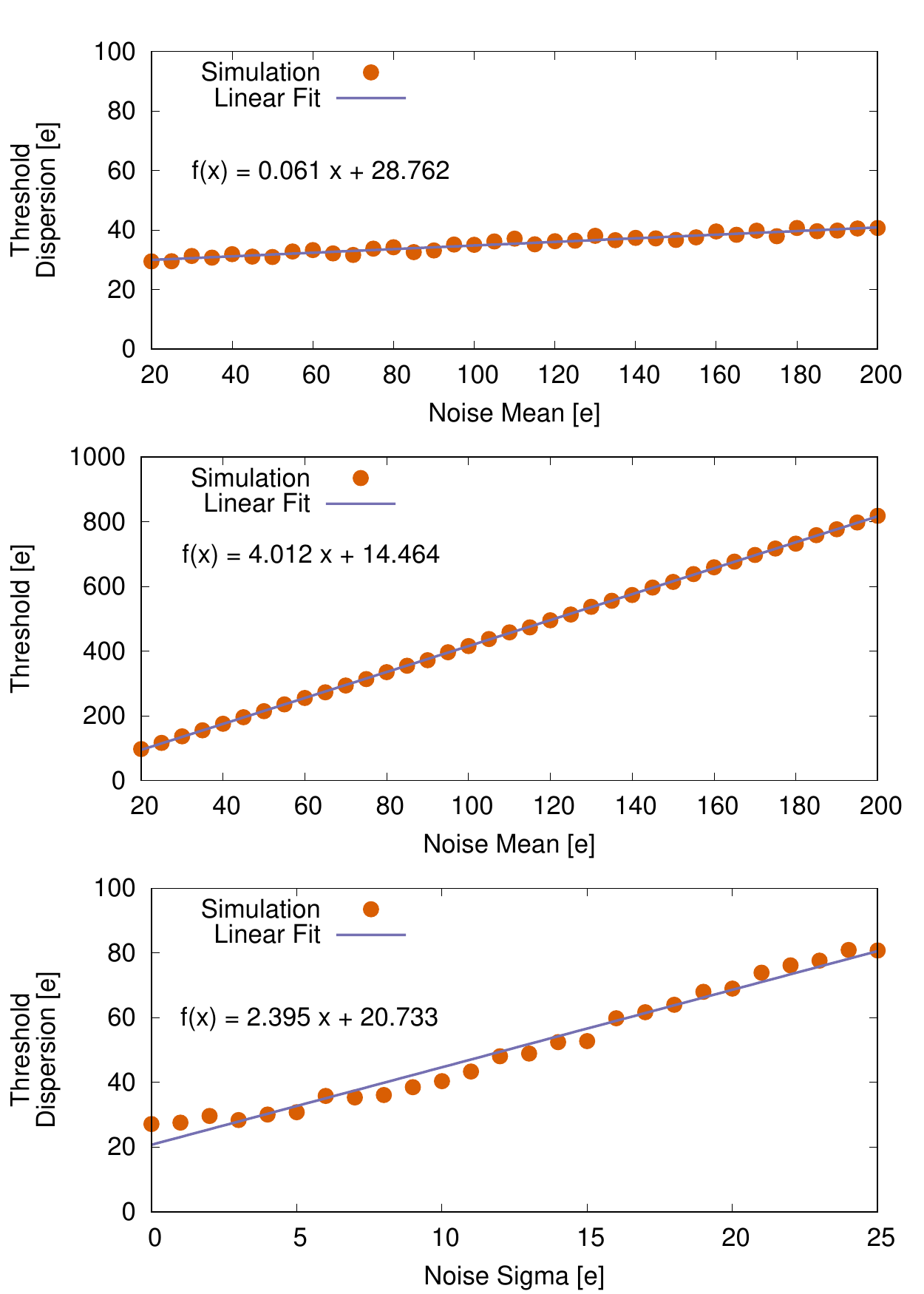}
	\caption{Behavior of threshold dispersion and threshold for different noise mean values. Particle hit rejection is 90\% and normalized drive clock frequency is 0.2.}
	\label{fig:noise}
\end{figure}

The width of the base threshold distribution does not influence the performance of the mechanism, unless the trim DAC range is too small to adjust for the width. But this is then equally true for manual tuning and needs to be addressed during design of the ASIC. The influence of the noise level is shown in Fig.~\ref{fig:noise}. An increase of the noise mean has no large effect on the threshold dispersion, but a higher mean noise will of course increase the threshold to achieve the same noise hit rate. The width of the noise however has a large effect on the threshold dispersion. As this mechanism tunes the threshold with reference to the noise level, there is clear correlation between the width of noise distribution and threshold dispersion. If the noise distribution is broader than the desired threshold dispersion, a traditional calibration via a test charge works better (recall that this method is built on the expectation that noise is well matched from pixel to pixel)..

\subsection{Trim DAC Step Size}
\begin{figure}[htbp]
	\centering
	\includegraphics[width=0.45\textwidth]{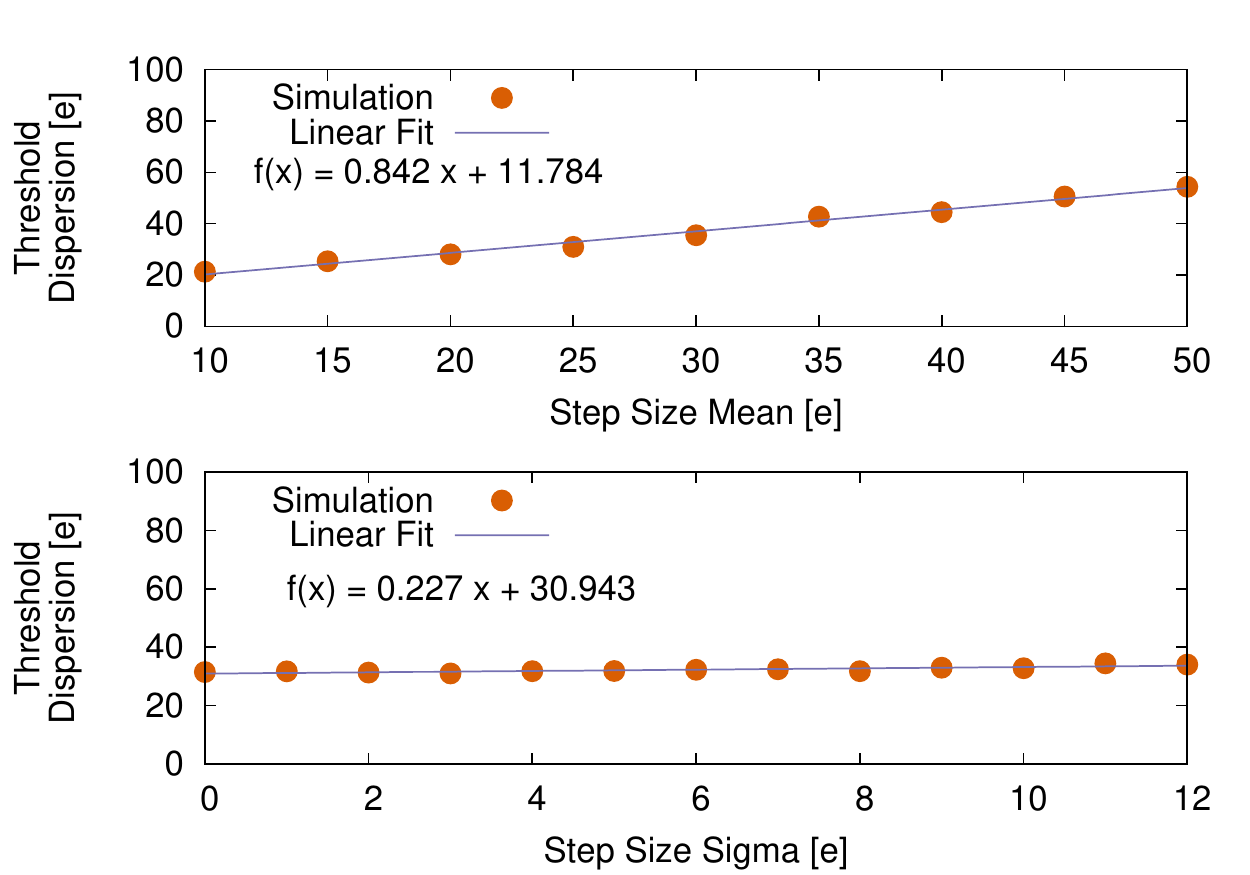}
	\caption{Behavior of threshold dispersion for different noise sigma values. Particle hit rejection is 90\% and normalized drive clock frequency is 0.2.}
	\label{fig:step}
\end{figure}

The influence of the trim DAC step size mean and sigma on the threshold dispersion are shown in Fig.~\ref{fig:step}. A finer step size leads to a lower threshold dispersion, but requires a larger trim DAC range. This is also true for the traditional tuning and should be optimized during the design of the ASIC. The width of the step size distribution has no or a negligible effect, which is to be expected as long as the DAC range is large enough.

\subsection{Noise Hit Acceptance}
\begin{figure}[h]
	\centering
	\includegraphics[width=0.45\textwidth]{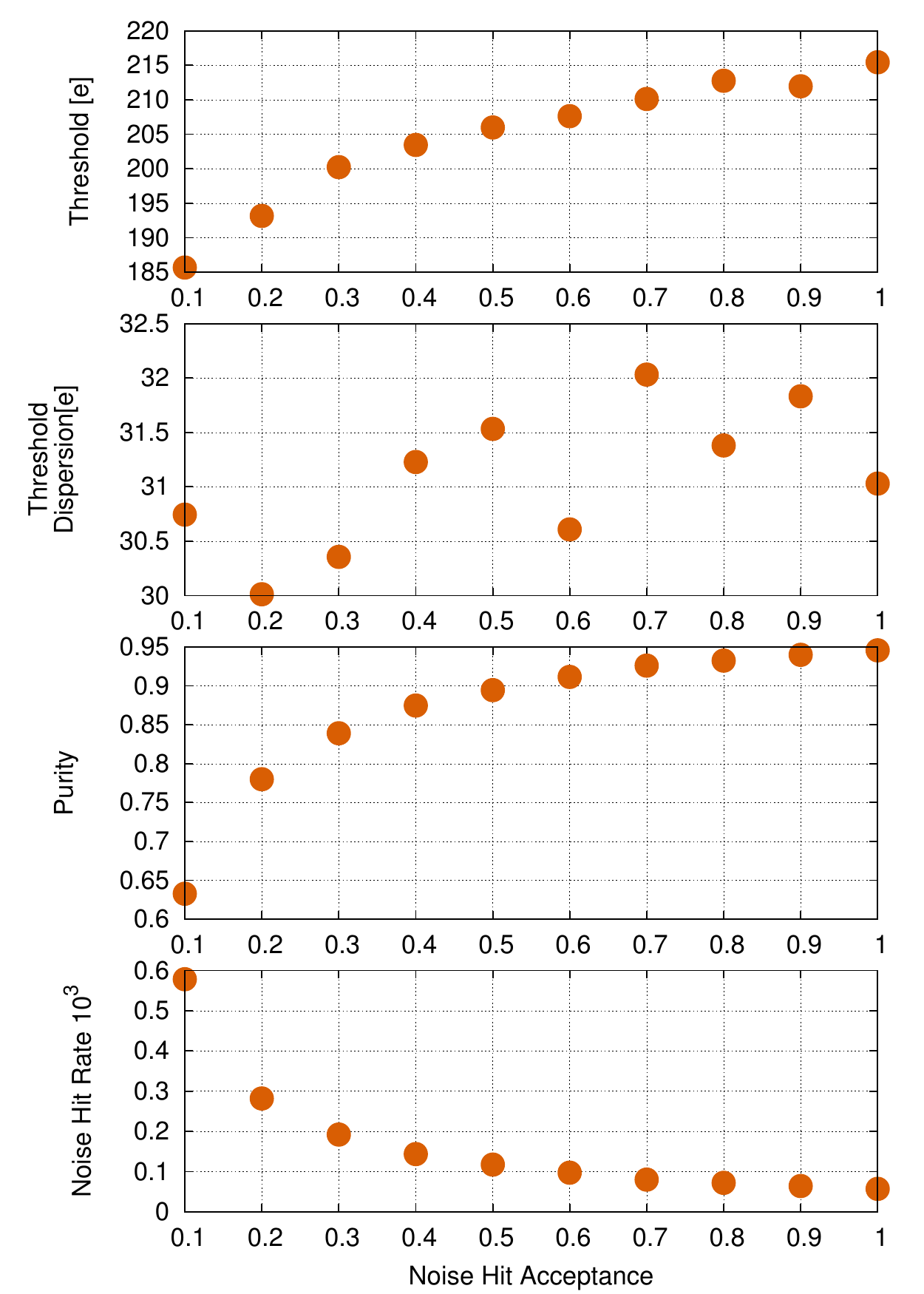}
	\caption{Behavior of mechanism for varying noise hit acceptance.}
	\label{fig:noisecut}
\end{figure}

So far we have assumed that the the filter accepts 100\% of all noise hit. The effect of less than perfect acceptance is shown in Fig.~\ref{fig:noisecut}. As a lower acceptance results in lower noise rate seen by the mechanism, it tunes to a lower threshold, which in return would result in a higher noise rate passing the threshold. As this behavior follows the  exponential dependence of noise rate on threshold, there is little to no effect down to an acceptance of even 70\%. The combinatorial logic in the filter should do better than this, as 99.9\% of all noise hits are isolated.

\section{Conclusion}
The proposed self-adjusting threshold mechanism for pixel detectors can help to overcome challenges in terms of stability and single event upset rate faced in modern detectors to be used in high radiation environments. For example, threshold dispersion increased of 5~e$^{-}$/C$^\circ$ and 160~e$^{-}$/kRad have been reported in \cite{fe65p2} for a prototype chip for the HL-LHC upgrade. Such large shifts would interfere with the operation of a readout chip and can be mitigated with the proposed mechanism.
Compared to it's analog implementation a digital circuit still allows the option of traditional configuration and tuning, making an inclusion in a new chip design very low risk. The analog implementation may be of particular interest for devices where the in-pixel transistor count must be minimized, 
such as monolithic active pixels. 
Simulation of the digital implementation shows that the achieved instantaneous threshold dispersion is on par with the one from traditional tuning, while the time variation is negligible, as long as the pixel to pixel noise variation is small relative to allowable threshold dispersion. It shows as well, 
that the mechanism works over a broad range of pixel parameters. 
In particular it can tolerate far less than perfect particle hit rejection of the filter logic.

\bibliographystyle{acm}
\bibliography{auto_tune}

\end{document}